# Research on Wikipedia Vandalism: a brief literature review

Jesús Tramullas
University of Zaragoza
C/ Pedro Cerbuna 12
50009 Zaragoza, Spain
tramullas@unizar.es

Piedad Garrido-Picazo
University of Zaragoza
Ciudad Escolar s/n
44003 Teruel, Spain
piedad@unizar.es

Ana I. Sánchez-Casabón
University of Zaragoza
C/ Pedro Cerbuna 12
50009 Zaragoza, Spain
asanchez@unizar.es

**ABSTRACT**
Research on vandalism in Wikipedia has been of interest for the last decade. This paper performs a literature review on the subject, with the goal of identifying the main research topics and approaches, methods and techniques used. 67 papers have been reviewed. Main topic is the detection of vandalism, although there is a increasing interest about content quality. The most commonly used technique is machine learning, based on feature analysis. It draws attention to the lack of research on information behavior of vandals.

## 1. Introduction.

A review of the bibliography published on a subject makes it possible to identify those aspects that are of particular interest for research. In recent years, Wikipedia has been the subject of various systematic bibliographic reviews [1]. The results of these reviews have been used to define and characterise Wikipedia as a subject of social and scientific/technical research and development [2, 3], and to identify significant topics or areas of research within it. One of the fields of research that appears related to the quality of the content of Wikipedia, and that is related to the collaborative editing process that supports it, is the study of vandalism. This user behaviour has been defined by Wikipedia itself as "... any addition, removal, or change of content, in a *deliberate* attempt to damage Wikipedia". As stated by Nielsen "Vandalism in Wikipedia presents usually not a problem: It will just be one more aspect to investigate" [4]. The interest in this aspect has been reflected in the holding of two *International Competitions on Wikipedia Vandalism Detection* in 2010 and 2011, in the context of the *Cross Language Evaluation Forum Conferences, CLEFs* [5, 6]. Two corpora of data have also been created for tests; the first of these is composed from articles of Wikipedia that have been edited [7], while the second has been created from Wikidata data editing [8].

Several systematic reviews have selected and studied the literature published about Wikipedia to identify research areas and fronts that have mentioned vandalism. Okoli [9] prepared a list of research topics without specifically citing vandalism. Martin [10] stressed the areas of quality (where he included vandalism), trust, semantic aspects, and governance and society. Jullien [11], from a perspective based on the study of the community, indicates motivation for participation, the interaction patterns and processes (with a minor mention of vandals) and quality assessment as areas of interest. Nielsen [4] classifies research into four major categories. These are identified as those that study Wikipedia (where he includes the study of vandalism), that use information from Wikipedia, that explore technical extensions (including those that could be used against vandalism), and those that use Wikipedia as a communication resource. Mesgari et al. [12], in the most recent review on the subject, grouped together papers in two large blocks, one relating to the quality of the content and the other to the size, without specifically dealing with vandalism.

The objective of this paper is to develop a systematic review of the academic literature published on vandalism on Wikipedia. The aim of this review is to: identify the applicable areas of knowledge in the subject (RQ1); the development over time of the number of publications (RQ2); the various research objectives proposed in the works consulted (RQ3); and the methods and techniques used (RQ4). We have not found any other similar paper in the available literature.

## 2. Method.

The method we used in this paper was a systematic review of the bibliography, which was applied to the literature about Wikipedia by Okoli and Schabram [13, 14]. The review that we have applied in this paper belongs to the category of stand-alone literature review, and its purpose is to systematically analyse the literature on the subject, without the use of primary data.

The specific techniques we followed has been previously applied and validated by Bar-Ilan and Aharony [2]. We did three searches for bibliographical references in the Web of  Science, Scopus and ACM DL databases, where we used the terms 'Wikipedia' and 'vandalism' in the title, key words and abstract fields. We did not search *Google Scholar* because it was not possible to limit the search terms to certain parts of the documents. The search carried out in *Web of Science* (titles and topics), on 1st February 2016, gave a result of 24 records. The search carried out in *Scopus* (title, key words and abstracts) on the same date, gave 92 results. The search also made on the same date in the *ACM Digital Library* (any field) gave 43 results.

The established protocol for the selection of documents has been the possibility to determine that their contents relate to a study into or analysis of vandalism on Wikipedia. We processed the obtained results in order to eliminate duplications, correct errors, filter out those documents that were not relevant to our search, and complete the basic data of each reference that had not been collected previously. We reviewed the content of each text. After this processing, the number of references obtained that we considered valid for the analysis was 67.

We reviewed each reference individually and we carried out qualitative data extraction and selection. We assigned an area of knowledge to each paper, depending on the affiliation of their authors and their contents. We used a controlled labelling system to describe the focus, objectives and the working method and we added the consolidated data to an RIS computer file for their processing using management software for bibliographical references, which for this paper was *Mendeley*, and to a CSV computer file, for their processing using data analysis applications. We released both files as *Open Data* so they could be used by other interested researchers. Finally, we published the references in specific groups on *Zotero. Mendeley, CiteUlike y Bibsonomy* (table 1).

| **Raw Data** | http://dx.doi.org/10.17632/dfxsb847cf.1 |
|---|---|
| ***Zotero*** | https://www.zotero.org/groups/wikipedia_vandalism_literature_review |
| ***Mendeley*** | https://www.mendeley.com/groups/8670151/wikipedia-vandalism-literature-review/papers/ |
| ***CiteUlike*** | http://www.citeulike.org/group/20074 |

**Table 1. Research data and references location.**

## 3. Results.

The basic quantitative analysis of the data obtained made it possible to obtain answers to the research questions we raised. The identified areas of knowledge (RQ1), depending on the affiliation of their authors, were related to computer science, humanities and social sciences, and biomedical sciences (figure 1).

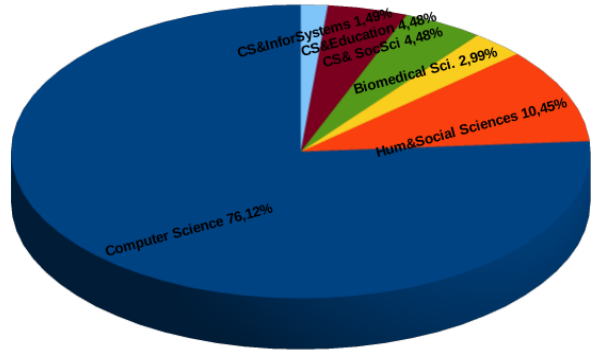

**Figure 1. Authorship by research area.**

Three quarters of the papers, 51 (76.12%) were written by researchers from the area of computer sciences. The humanities and social sciences area contributed 7 (10.44%). The publications from the field of biomedical sciences were very low, 2 (2.99%). The number of papers that we identified as a result of collaboration between the various research areas was 7 (10.45%), in which computer science participated jointly with social sciences (3 papers), education (3 papers) and information systems (1 paper).

The time period when the analysed papers were published was between 2007 and 2015 (RQ2). Most of the publications occurred in 2010 (15) and 2011 (11), and were related to the holding of the International Competitions on Wikipedia Vandalism Detection. This number decreased later, with there being 5 and 7 publications identified in 2014 and 2015 respectively (figure 2).

The objectives of the various papers reviewed demonstrate the concentration on just a few research topics (RQ3, figure 3). The detection of vandalism is defined as the most widely sought objective: 40 papers were devoted to this (59.7%). This is followed by the less important quality control of Wikipedia content (8, 11.94%), analysis of the textual content (4, 5.97%) and prevention (2, 2.99%). Objectives such as the analysis of reliability, the elimination or the automatic blocking of acts of vandalism, the study of editors and of revertings, or the forecasting of possible acts of vandalism appear with just one mention. Only two papers (2.99%) can be considered as approaches to the study of the information behaviour that underlies vandalism. Regarding development over time, the majority of the papers on detection appear up to 2013, while in 2014 and 2015 they are equalled by those focussing on content quality.

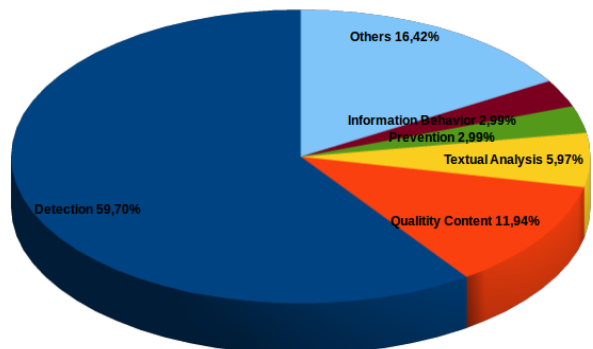

**Figure 3. Research Objectives.**

In order to achieve the research objectives, the technique that was most widely used was classification by machine learning, based on feature analysis, which was identified in 34 papers (RQ4). In contrast, other papers state that quantitative and metric techniques were applied to carry out feature analysis, but no further details are given. Study into the reputation of editors and articles was also an object of interest, appearing in 6 papers both as an objective and method. In 4 papers, space-time metadata were analysed using simulation techniques; in other cases, multi-dimensional analysis and probabilistic modelling or analysis, or simply the use of classifiers, were applied. We should mention the approaches based on qualitative studies, which represent 7 papers (6 of them from humanities and social sciences), and which show a classic research methodology in these disciplines (figure 4).

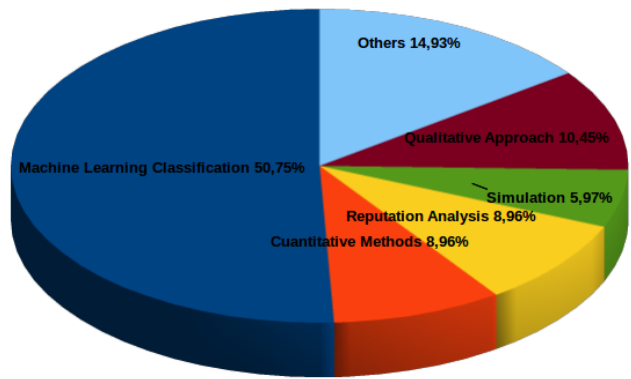

**Figure 4. Technical Approaches.**

If we focus on software production, the papers reviewed show very few products: we identified *WikiTrust, iChase, Stiki, VandalSense and Illiterate Editor,* of which only the codes from *Stiki* [15] and *WikiTrust* [16] are available, although the latter has been discontinued.

## 4. Discussion.

The first question to be dealt with is the validity of the sample. Of the 92 papers recovered, we considered 67 to be relevant for the study. This means that we rejected almost 30% of the papers, despite them fulfilling the search terms, because these were not relevant to our research. Given the fact that titles, abstracts and key words are drafted by researchers themselves, this high rejection rate indicates that the selection of papers for a literature review through a traditional method based solely on search results should be complemented by a qualitative review of the papers, as the information provided by the authors, which appears in the databases of reference, is not completely reliable.

- RQ1: Our analysis of the research areas demonstrates the preponderance of computer science in the study of vandalism on Wikipedia. Over 75% of the papers come from this scientific field. The papers from the field of humanities and social sciences only reach 15%. We should also mention the low level of collaboration between researchers from the various research areas. This results in a dominance of papers on partial and specific aspects of vandalism, and does not make it possible to find integrative approaches to the problem.

- RQ2: If we contextualise the number and the development over time of the papers within the general context of publications outlined in [2], we have to state that the attention devoted to vandalism on Wikipedia is low. It does not appear to be a priority area of research, either in the technological or social field. Regardless of the low number, the development over time appears to support the main guidelines established in the cited paper.

- RQ3: The main objective of the published papers is the detection of vandalism, and the improvement of the efficiency and effectiveness of the methods used. We have identified the authors of these papers as coming from the field of computer science. Other objectives, such as the quality of information and content analysis are of secondary importance. We should emphasise the low number of papers that study the informational behaviour of users and editors regarding vandalism. If we take into consideration the fact that vandalism is a human activity, and that the Wikipedia data dumps make it possible to study, both qualitatively and quantitatively, user behaviour, then this vacuum is surprising. The precise objective of the papers on the reputation of reviewed editors is vandalism detection, ahead of the understanding of behaviour. Neither have they dealt with aspects such as edit conflicts, the lifecycle of articles or discussions on user pages in any great detail.

- RQ4: The most used method has been the use of classifiers, in machine learning processes, for the detection of acts of vandalism, against a previously established corpus [17]. The analysis that the researchers carry out is the same as outlined by Adler et al [18], and relates to one of the four basic computational approaches: language characteristics, textual content characteristics, metadata relating to publications and the reputation of editors. It would be helpful if the papers indicated the type or types of acts of vandalism that they study, given that this can relate to deletions, misinformation, offensive content, spam, non sequiturs, etc. Unlike the majority of papers from the field of computer science and the use of quantitative methods, the papers that chose qualitative methodologies are low in number and come almost exclusively from humanities and social science.

Finally, we should stress the almost complete lack of impact that research appears to have had on the development of software tools. Only *Stiki* is available and includes maintenance.

## 5. Conclusion and future work.

Our discussion and review of the results obtained enable us to state that research into vandalism does not occupy a central position in research into Wikipedia. The main objective has been to develop better methods that will make the automatic detection of acts of vandalism possible, to then expand the focus towards the analysis and classification of reputations and quality assurance of the textual content.

However, the research carried out so far lacks a sufficient number of fundamental studies into the informational behaviour of vandals. Research into the phenomenon has not been carried out from an integrated perspective that combines computational methods with the necessary sociotechnical research. Vandalism of the content of articles, as human behaviour, is not the only conflict that occurs on Wikipedia: edit conflicts, arguments, user discussion pages and the *village pump* threads of the various Wikipedias are all environments where we can find information for research into vandalism on Wikipedia. An example of this would be trolling behaviour. A more integrated focus cannot deal with the problem solely through approaches based on the classification of edited content as vandalism or not.

Detection and elimination of vandalism would benefit from a broader understanding of the problem and, in turn, should be included within the general context of information quality on Wikipedia. This context should include research into the reputation of editors. However, this approach should not be exclusively computational: Wikipedia is a community with principles and unwritten social rules that should be taken into consideration when studying it.

Another aspect to be developed would be the conversion of research results into software tools that would be useful for the community of Wikipedia editors and users, which is an aspect that until now has hardly developed. New developments in active Wikipedia tools, like recent *Objective Revision Evaluation Service* [19] a set of APIs designed to provide support for quality control tools, are oriented to support the activity of editors and librarians.

Finally, this paper could be developed by carrying out an analysis of the concept of quality on Wikipedia, and by understanding that research into vandalism is an aspect within it. Detailed bibliometric analysis could reveal new connections between groups of researchers, research topics, lines of influence and guidelines for the use of information between them.

# 6. References.